\documentstyle{article}
\begin{document}
\bibliographystyle{stand}

\title{
{\rm
\rightline{hep-ph/9707468}
\ \\
\ \\}
(SEMI)CLASSICAL MOTION IN FIELDS
OF AHARONOV-BOHM AND AHARONOV-CASHER}

\author{Ya.I.Azimov$\,{}^a$, R.M.Ryndin$\,{}^b$\\
Petersburg Nuclear Physics Institute\\
Gatchina, St.Petersburg, 188350, Russia\\
{\it e-mail: ${}^a\,$azimov@pa1400.spb.edu,}\\
{\it $~~~~~~~~~{}^b\,$ryndin@thd.pnpi.spb.ru}}
\date{}
\maketitle

\begin{abstract}
Particle motion in the fields of Aharonov-Bohm and Aharonov-Casher is
considered in framework of the classical theory to reveal conditions
admitting duality of the two configurations. Important role of
orientation of the magnetic dipole moment is demonstrated. Duality
becomes totally destroyed by addition of electric dipole and/or higher
multipole moments. Correspondence between quantum and classical
considerations is also discussed.


\end{abstract}

\vspace{1cm}

1. The Aharonov-Bohm effect (ABE)~\cite{AB}, the first example of
purely topological effects closely related to the gauge character of
electromagnetic field, continues to be of scientific and applied
interest. Since its discovery the effect has been intensively studied
in many details, but it was only rather recently that another
topological effect in electromagnetic fields, analogous to ABE, was
suggested, namely, the Aharonov-Casher effect (ACE) in purely electric
field~\cite{AC}.

The question of existence of ACE initiated a theoretical discussion
(see refs.\cite{B,Y,G}). It had not been concluded when the neutron
beam measurements~\cite{C} confirmed ACE both qualitatively and with
rather good quantitative precision. Even more precise value for the
Aharonov-Casher phase was obtained with molecular beams in the
homogeneous electric field~\cite {H1,H2}. And the theoretical
discussion was closed by the paper of Hagen~\cite{Hag} who demonstrated
exact duality of the Dirac wave functions in the fields of
Aharonov-Bohm (AB) and Aharonov-Casher (AC).

The Hagen approach was later generalized for an arbitrary spin $s$,
with the result~\cite{AR}: the AB and AC configurations are mutually
`dual only at the extreme spin projections, $\pm s$, normal to
the field plane. Such result in the 2-dimensional problem seems to
present a paradox:  according to the group theory the total angular
momentum should be inessential, its role should be played by the normal
projection. For example, one could expect that the particle motion at
spin 3/2 and its normal projection 1/2 should be equivalent to the
motion at the same normal projection, but with spin 1/2.  However, the
AB and AC configurations are dual for the latter case and not dual for
the former one.

In the present note we show that purely classical consideration of
particles in the fields of AB and AC reveals physical reasons for
such special properties of the extreme spin projections. We can also
study influence of higher multipole moments, above the charge and
magnetic moment.

2. Begin with consideration of a purely classical particle in electric
and magnetic fields. Comparison of papers~\cite{B,Y,G} to each other
and to earlier studies~\cite{P,Fr} reveals absence of the generally
accepted relativistic Lagrangian $L_{int}$ for interaction of
point-like multipoles with external fields. We construct it from two
requirements:  \begin{itemize} \item In the rest-frame of the particle
(at ${\bf v}=0$) $L_{ int}$ equals to the corresponding potential
energy with sign reversed; \item $L_{int}$ transforms as
$dS/dt$ with $S$ being the invariant action, so it is an invariant
multiplied by \mbox{$(1-v^2)^{1/2}$}.  \end{itemize}
In this way we obtain, e.g., the following interaction Lagrangians for
the point charge and the point-like electric and magnetic dipoles (such
procedure gives, of course, the standard form to $L^{(ch)}$):
\begin{equation} L^{(ch)}({\bf x},t)=-q\phi({\bf x},t)+
q{\bf v}\cdot{\bf A(x},t)\ , \;\; \;\; \end{equation}
\begin{equation}
L^{(ed)}({\bf x},t) = {\bf d}\cdot{\bf E(x},t)+{\bf d}\cdot [{\bf
vB(x},t)]\ , \; \end{equation}
\begin{equation} L^{(md)}({\bf
x},t)={\bf m}\cdot {\bf B(x},t)-{\bf m}\cdot [{\bf vE(x},t)]\ .
\end{equation}
There exists a tradition (see, e.g., refs.\cite{P,Fr}) to describe
electric and magnetic dipoles by a combined antisymmetric tensor
(similar to the field tensor {\bf(E,B)}). However, they play very
different roles for elementary particles (e.g., ${\bf d}\neq0$ leads
to $P$ and $T$ violation). So we use two independent tensors, each one
having a special form in the particle rest-frame: $(-{\bf d}_0,0),
(0,{\bf m}_0)$. In an arbitrary system these tensors are $(-{\bf d}
(1-v^2)^{-1/2},[{\bf dv}](1-v^2)^{-1/2})$ and $([{\bf mv}](1-v^2)^{-1/2},
{\bf m}(1- v^2)^{-1/2})$ if the particle moves with the velocity
{\bf v}.

Correct relativistic transformation of $L_{int}$ is provided by $q=\mbox
{inv}$, while {\bf d} and {\bf m} should change. Transition from the
rest-frame to a moving one transforms them just in the same way as the
radius vector between two space points: longitudinal components undergo
the Lorentz contraction, transversal ones do not change. Nevertheless,
we do not vary them with {\bf v}: since dipole moments of a classical
particle may change with time (in both value and direction) even in
its rest-frame, we assume them to depend directly on time but not on
velocity which is time-dependent itself.

The above interaction Lagrangians explicitly depend on velocity, so the
particle canonical momentum contains both the familiar contribution,
proportional to the vector potential, and other contributions,
proportional to other multipoles. $L^{(ch)},\ L^{(ed)}$ and $L^{(md)}$
produce additional terms
\begin{equation} \delta{\bf p}=q_e{\bf
A(x},t)-[{\bf dB(x},t)]+[ {\bf mE(x},t)]\ .  \end{equation}
The last two terms are sometimes considered as "hidden" momenta, but
these extra terms are really induced by the explicit dependence of
interaction on velocity.

Every multipole generates the corresponding forces of the particle
interaction with fields. For {\bf d} and {\bf m} the forces may be
presented as ${\bf f}={\bf f}_{(0)}+\delta{\bf f}$, where
\begin{equation} {\bf f}^{(ed)}_{(0)}=d_i\partial_i{\bf E}+ d_i
[{\bf v}\partial_i{\bf B}] + [\dot{\bf d}{\bf B }]\ ,\;\;
\end{equation}
\begin{equation} {\bf f}^{(md)}_{(0)}= m_i\partial_i{\bf B}- m_i
[{\bf v}\partial_i{\bf E}] - [\dot{\bf m}{\bf E}]\ .
\end{equation}
Because of the Maxwell equations, the extra term $\delta{\bf f}^{(ed)}$
disappears while $\delta{\bf f}^{(md)}$ appears to be expressed through
charge and current densities produced by the particle itself. So (if
not vanishing) it corresponds to the particle self-interaction which
influence should be included into the particle mass and/or other
parameters, but should not arise in its equation of motion under
external forces.  Note that the forces ${\bf f}_{(0)}$ may be obtained
by describing dipoles as limiting systems of two separated
opposite charges, electric $\pm q_e$  or magnetic $\pm q_m$ (if ithey
existed), each one subjected to the electric or magnetic Lorentz force
\begin{equation}
{\bf f}^{(e)}= q_e{\bf E} + q_e[{\bf vB}]\ ,\;\;\; {\bf f}^{(m)}=
q_m{\bf B} - q_m[{\bf vE}]\ .  \end{equation}
${\bf f}^{(e)}$ has the familiar form and arises directly from
$L^{(ch)}$; the both forces may be obtained by the Lorentz transformation
of the force acting on a motionless charge, respectively electric or
magnetic. Strange enough, none of papers~\cite{AC,B,Y,G} on ACE gives the
total classical force (6), though each of them correctly presents its
separate terms.

There is one more point which has not been discussed at all in
connection with ACE. A particle bearing dipole moments (or higher
multipoles), even being point-like, should be considered as anisotropic.
Its orientation produces new degrees of freedom. The particle can have
non-vanishing angular momentum, it can be subjected to torques.
Expressions for the torques may be obtained from $L_{int}$ by varying
the particle orientation. For dipole moments the torques are
\begin{equation} {\bf
M}^{(ed)}=[{\bf dE}]+[{\bf d[vB}]]\ ,\; \end{equation} \begin{equation}
{\bf M}^{(md)}= [{\bf mB}]-[{\bf m[vE}]]\ . \end{equation}
We shall see in what follows that torques are very important for
understanding properties of the particle motion.

In a similar way one can also find Lagrangians, forces and torques for
particles with any higher multipoles. We do not give them here
explicitly, but will have them in mind to consider the contribution of
higher multipoles.

Let us briefly discuss how the motion influences interaction of
multipoles with external fields. For $L_{int}$ and {\bf M} it
leads only to a formal change of real fields by effective ones
${\bf E}'={\bf E}+[{\bf vB}],\,\, {\bf B}'={\bf B}-[{\bf vE}]$.
But the change is not the only physical effect, as can be seen from
additional contributions to the canonical momentum and force.

3. Let us apply Lagrangians (1)-(3) to the classical motion of a
particle in the Aharonov-Bohm (AB) field, i.e. in the field of the
infinite thin solenoid along the $z$-axis. Field strengths vanish
around the solenoid; the motion there looks free, without any forces
and torques. However, {\bf p} contains an extra term $\delta{\bf p(x})$
with  $\delta p_z=0$.

The situation becomes less trivial if we consider the thin solenoid as
the limit of a solenoid with a finite radius. For definiteness we
suggest the following structure of the magnetic field strength: from
the $z$-axis up to some radius the field is homogeneous and directed
along the $z$-axis; in the transient region the field conserves its
direction and diminishes its strength down to zero; the field strength
(but not potential!) is absent everywhere outside. We also assume that
initially the particle moves in the $(x,y)$ plane orthogonal to the
solenoid.

If the particle in its motion does not go through the strength region
then no force and no torque arise; the classical motion stays free. In
the strength region, as is well known, the field produces the force
${\bf f}^{(e)}$. It acts on the particle charge and tends to curve its
trajectory without driving it out of the $(x,y)$-plane. Neither
the particle orientation changes.

Let us consider the role of magnetic moment. If the dipole is normal to
the motion plane (i.e. along the $z$-axis) then, according to (6) and
(9), it is not subjected to any torque or additional force. The
particle orientation conserves.

The situation is quite different if {\bf m} has non-vanishing projection
onto the motion plane. Because of the torque ${\bf M}^{(md)}$ the
dipole {\bf m} precesses around the field direction. Precession does
not produce any additional force in the homogeneous region. However,
in the transient region it does generate the force ${\bf f}^{(md)}$
which has a non-vanishing $z$-component and draws the motion out of
the $(x,y)$-plane.  The sign of $f^{(md)}_z$ may change in the course
of precession, so the motion oscillates around its initial plane. Thus,
the particle leaves the field region with, generally, non-zero $v_z$.
Its sign and value depend on both the field structure in the transient
region and the initial orientation of {\bf m}.  Projection of {\bf m}
onto the {\bf B}-direction does not change.

The motion becomes much more complicated if the particle, in addition
to the charge and magnetic moment, has also an electric dipole moment
(EDM) or higher multipoles. In such a case the precession axis itself
changes its direction, and the particle orientation evolves in a rather
complicated way. There are additional forces and torques acting in both
transient and homogeneous regions, at any initial particle orientation.
As a result, after going through the field  the particle not only gets
out of plane, but also changes its {\bf m}-projection onto {\bf B}.

So, we see that a classical particle in the field of a long solenoid
moves differently in the presence, in addition to the charge, of some
other multipoles (even magnetic moment). Resulting motion is very
sensitive to the field structure, and, hence, the problem of the
infinitely long and thin solenoid may essentially depend on the
limiting procedure. Only for the particle with charge and magnetic
moment, exactly along the field direction, the motion has "canonical"
properties.

4. Now we consider a classical particle in the Aharonov-Casher (AC)
field, i.e., in the field of a uniformly charged straight thin thread
along the $z$-axis. Its electric strength is perpendicular to the
thread and depends only on two coordinates $x$ and $y$. We assume the
particle to be neutral, but having a magnetic moment and, may be, some
other multipoles. Initially the particle moves in the $(x,y)$-plane.

Let us begin with {\bf m} oriented along the $z$-axis. Then eq.(9)
shows that torques are absent, and the particle orientation does not
change. Furthermore, $\dot{\bf m}=0,\,\,m_x=m_y=0,$ and, according to
eq.(6), forces are also absent. The particle motion looks to be free,
but its canonical momentum, according to eq.(4), differs from the free
value $m{\bf v}(1-v^2)^{-1/2}$. The extra term $\delta{\bf p}$ has the
same structure as for the charged particle in the AB field outside the
solenoid (in particular, $\delta p_z=0$). Therefore, all the classical
description of this case is similar to the case of a charged particle
in the AB field.

If {\bf m} has non-vanishing projection onto the $(x,y)$-plane, then
$\delta p_z\neq 0$. This means that $\delta{\bf p}$ is
definitely different here from the extra term in the AB case. Such
motion generates the torque which compels {\bf m} to precess around the
axis $[{\bf vE}]$, initially parallel to the $z$-direction. Therefore,
non-vanishing $\dot{\bf m}$ arises, initially in the $(x,y)$-plane.
Note that $\dot{\bf m}$ and {\bf m} are orthogonal to each other, so
the absolute value $|{\bf m}|$ does not change. Now, non-zero
$\dot{\bf m}$ and projections of {\bf m}  onto the $(x, y)$-plane
generate, according to (6), a force (with non-zero $z$-component) which
changes the {\bf v}-direction and even draws the motion out of the
plane.

If one compares such classical motion in the AC configuration to the
classical motion of a charged particle in the AB field with the same
initial orientation of {\bf m}, then an essential difference can be
seen.  The AB motion is also subjected to forces and torques violating
its plane character, but only when propagating through the confined
region of the solenoid itself. In the AC field the forces and torques
act over the whole space. Thus, the AC motion with oblique orientation
of {\bf m} can never be free and never is similar to the AB motion. The
magnetic moment {\bf m} conserves its absolute value, but changes its
orientation in a rather complicated way: it precesses around the axis
$[{\bf vE}]$ which itself changes its direction.

Nearly of the same character is the motion when the particle carries
also EDM or higher multipoles. But time evolution of {\bf m} and {\bf
v} becomes more complicated. One of the reasons is that addition of any
other multipoles to the magnetic moment induces forces and torques
which affect the motion in the AC field even for the initial {\bf
m}-orientation along the $z$-axis. Therefore, possibility of duality
between the classical AB and AC motions becomes totally destroyed.

5. In conclusion, we briefly discuss correspondence between classical
and quantum descriptions of multipole interactions for a moving
particle. We have shown in the preceding Sections that similarity
between the classical AB and AC motions is possible only if the
magnetic moment was initially oriented exactly along the $z$-axis. Such
a conclusion is just the same as the result of the quantum
consideration~\cite{AR} and reveals its physical reason.

Furthermore, the classical study has shown that addition of EDM or
higher multipoles totally prevents possible AB--AC duality. Leaving
apart detailed quantum analysis of such a case we consider here only
briefly how the multipoles influence the particle wave function.

In the AB field the particle carrying only charge and magnetic moment
conserves its spin $z$-projection~\cite{Hag,AR}. This property for a
Dirac particle in the AB field is expressed by commutation of $\gamma_3
\gamma_5$ with the Dirac operator, if there is only magnetic field
directed along the $z$-axis and the wave function depends only on $x$
and $y$. Presence of EDM induces interaction proportional to
$\sigma_{\mu\nu}\gamma_5 F^{\mu\nu}$ ($P$ and $T$ violation does not
matter here). If the field consists only of $B_z$, we have the operator
contribution $\sigma_{12}\gamma_5$ which does not commute with
$\gamma_3\gamma_5$, and the spin orientation cannot conserve. EDM has
similar effect in the AC configuration, since it induces the terms
$\sigma_{0k}\gamma_5$ with $k=1,2$ in the Dirac operator.  However, the
quantitative influence of EDM on the AB and AC wave functions is
different, both because of different properties of fields {\bf B} and
{\bf E} and due to their different space distributions.  So, the AB--AC
duality is excluded by the presence of EDM.

Higher multipoles require a higher spin and complicate relativistic
consideration (see, e.g.,~\cite{AR}). But for the qualitative
understanding we may stick to the quadrupole moment in nonrelativistic
approximation.  Operator of the quadrupole moment, both electric and
magnetic, is proportional to $S_j S_k + S_k S_j -\frac {2}{3} {\bf
S}^2\delta_{jk},$ where {\bf S} is the spin operator. It is easy to
check that in both configurations, AB and AC, the interaction contains
terms not commuting with $S_3$ and violating the spin $z$-component
conservation. The AB--AC duality is again impossible. So, the quantum
description of the role of various multipoles also agrees with
classical one, at least qualitatively.

One of interesting classical results is the inevitable non-plane
character of motion if the magnetic moment was initially deflected from
the $z$-axis (or in the presence of EDM or higher multipoles).
Meanwhile, the quantum wave function may still depend on
$x$ and $y$ only which is usually thought to indicate plane motion.
However, really it is not so. At the space infinity, where {\bf E}
vanishes, the canonical momentum coincides with $m{\bf
v}(1-v^2)^{-1/2}$. Let us assume that $v_z=0$ there and so {\bf p} lies
in the $(x,y)$-plane.  In the AC motion $p_z$ conserves and {\bf p}
remains in the plane. That is why the wave function depends only on two
variables. But, if {\bf m} was deflected from the $z$-axis, {\bf v}
deviates from {\bf p} (see eq.(4)) and aquires non-zero $z$-component.
Only at $|m_z|=|{\bf m}|$ the problem degenerates into the really plane
one. Detailed comparison of quantum and classical considerations here
may require to define more precisely what is the velocity (and, m.b.,
coordinate as well) for a quantum relativistic particle in the external
field.

As a result of classical consideration and its comparison to quantum
one we can note an interesting difference between ABE and ACE. In the
field of a thin solenoid the wave function of the charged particle,
while going around the solenoid, produces the known geometrical phase
independently of the form of the solenoid and orientation of the particle
magnetic moment (the solenoid is straight only for simplicity; particle
orientation is not important since the field strength vanishes around
the solenoid). On the contrary, orientation is very essential in the
field of a thin charged thread: even for the straight thread the
geometrical phase arises only for two extreme orientations of the
magnetic moment. But if the thread is curved, the magnetic
moment projection onto the thread direction cannot be fixed, and going
around the thread is not related to the geometrical phase. Therefore,
ACE disappears if the charged thread is bent, while the ABE is just a
direct consequence of the presence of the solenoid, independently of
its form. In that sense, ACE is not a real topological effect.

This work has been supported by the grant RFBR 96-02-18630. The authors
are grateful to A.Vainshtein and V.Zelevinsky for stimulating
discussions.

\end{document}